\documentclass[ALICE,manyauthors]{cernphprep}
\usepackage[comma,square,numbers,sort&compress]{natbib}
\usepackage{hyperref}
\usepackage{lineno}
\usepackage{xspace}
\usepackage{xcolor}
\usepackage{orcidlink}
\usepackage[T1]{fontenc}
\begin{document}
%

\newcommand{\pp}           {pp\xspace}
\newcommand{\ppbar}        {\mbox{$\mathrm {p\overline{p}}$}\xspace}
\newcommand{\XeXe}         {\mbox{Xe--Xe}\xspace}
\newcommand{\PbPb}         {\mbox{Pb--Pb}\xspace}
\newcommand{\pA}           {\mbox{pA}\xspace}
\newcommand{\pPb}          {\mbox{p--Pb}\xspace}
\newcommand{\AuAu}         {\mbox{Au--Au}\xspace}
\newcommand{\dAu}          {\mbox{d--Au}\xspace}

\newcommand{\s}            {\ensuremath{\sqrt{s}}\xspace}
\newcommand{\snn}          {\ensuremath{\sqrt{s_{\mathrm{NN}}}}\xspace}
\newcommand{\pt}           {\ensuremath{p_{\rm T}}\xspace}
\newcommand{\meanpt}       {$\langle p_{\mathrm{T}}\rangle$\xspace}
\newcommand{\ycms}         {\ensuremath{y_{\rm CMS}}\xspace}
\newcommand{\ylab}         {\ensuremath{y_{\rm lab}}\xspace}
\newcommand{\etarange}[1]  {\mbox{$\left | \eta \right |~<~#1$}}
\newcommand{\yrange}[1]    {\mbox{$\left | y \right |~<~#1$}}
\newcommand{\dndy}         {\ensuremath{\mathrm{d}N_\mathrm{ch}/\mathrm{d}y}\xspace}
\newcommand{\dndeta}       {\ensuremath{\mathrm{d}N_\mathrm{ch}/\mathrm{d}\eta}\xspace}
\newcommand{\avdndeta}     {\ensuremath{\langle\dndeta\rangle}\xspace}
\newcommand{\dNdy}         {\ensuremath{\mathrm{d}N_\mathrm{ch}/\mathrm{d}y}\xspace}
\newcommand{\Npart}        {\ensuremath{N_\mathrm{part}}\xspace}
\newcommand{\Ncoll}        {\ensuremath{N_\mathrm{coll}}\xspace}
\newcommand{\dEdx}         {\ensuremath{\textrm{d}E/\textrm{d}x}\xspace}
\newcommand{\RpPb}         {\ensuremath{R_{\rm pPb}}\xspace}

\newcommand{\nineH}        {$\sqrt{s}~=~0.9$~Te\kern-.1emV\xspace}
\newcommand{\seven}        {$\sqrt{s}~=~7$~Te\kern-.1emV\xspace}
\newcommand{\twoH}         {$\sqrt{s}~=~0.2$~Te\kern-.1emV\xspace}
\newcommand{\twosevensix}  {$\sqrt{s}~=~2.76$~Te\kern-.1emV\xspace}
\newcommand{\five}         {$\sqrt{s}~=~5.02$~Te\kern-.1emV\xspace}
\newcommand{\twosevensixnn}{$\sqrt{s_{\mathrm{NN}}}~=~2.76$~Te\kern-.1emV\xspace}
\newcommand{\fivenn}       {$\sqrt{s_{\mathrm{NN}}}~=~5.02$~Te\kern-.1emV\xspace}
\newcommand{\LT}           {L{\'e}vy-Tsallis\xspace}
\newcommand{\GeVc}         {Ge\kern-.1emV/$c$\xspace}
\newcommand{\MeVc}         {Me\kern-.1emV/$c$\xspace}
\newcommand{\TeV}          {Te\kern-.1emV\xspace}
\newcommand{\GeV}          {Ge\kern-.1emV\xspace}
\newcommand{\MeV}          {Me\kern-.1emV\xspace}
\newcommand{\GeVmass}      {Ge\kern-.2emV/$c^2$\xspace}
\newcommand{\MeVmass}      {Me\kern-.2emV/$c^2$\xspace}
\newcommand{\lumi}         {\ensuremath{\mathcal{L}}\xspace}

\newcommand{\ITS}          {\rm{ITS}\xspace}
\newcommand{\TOF}          {\rm{TOF}\xspace}
\newcommand{\ZDC}          {\rm{ZDC}\xspace}
\newcommand{\ZDCs}         {\rm{ZDCs}\xspace}
\newcommand{\ZNA}          {\rm{ZNA}\xspace}
\newcommand{\ZNC}          {\rm{ZNC}\xspace}
\newcommand{\SPD}          {\rm{SPD}\xspace}
\newcommand{\SDD}          {\rm{SDD}\xspace}
\newcommand{\SSD}          {\rm{SSD}\xspace}
\newcommand{\TPC}          {\rm{TPC}\xspace}
\newcommand{\TRD}          {\rm{TRD}\xspace}
\newcommand{\VZERO}        {\rm{V0}\xspace}
\newcommand{\VZEROA}       {\rm{V0A}\xspace}
\newcommand{\VZEROC}       {\rm{V0C}\xspace}
\newcommand{\Vdecay} 	   {\ensuremath{V^{0}}\xspace}

\newcommand{\ee}           {\ensuremath{e^{+}e^{-}}} 
\newcommand{\pip}          {\ensuremath{\pi^{+}}\xspace}
\newcommand{\pim}          {\ensuremath{\pi^{-}}\xspace}
\newcommand{\kap}          {\ensuremath{\rm{K}^{+}}\xspace}
\newcommand{\kam}          {\ensuremath{\rm{K}^{-}}\xspace}
\newcommand{\pbar}         {\ensuremath{\rm\overline{p}}\xspace}
\newcommand{\kzero}        {\ensuremath{{\rm K}^{0}_{\rm{S}}}\xspace}
\newcommand{\lmb}          {\ensuremath{\Lambda}\xspace}
\newcommand{\almb}         {\ensuremath{\overline{\Lambda}}\xspace}
\newcommand{\Om}           {\ensuremath{\Omega^-}\xspace}
\newcommand{\Mo}           {\ensuremath{\overline{\Omega}^+}\xspace}
\newcommand{\X}            {\ensuremath{\Xi^-}\xspace}
\newcommand{\Ix}           {\ensuremath{\overline{\Xi}^+}\xspace}
\newcommand{\Xis}          {\ensuremath{\Xi^{\pm}}\xspace}
\newcommand{\Oms}          {\ensuremath{\Omega^{\pm}}\xspace}
\newcommand{\degree}       {\ensuremath{^{\rm o}}\xspace}

\begin{titlepage}
\PHyear{2024}        
\PHnumber{344}       
\PHdate{18 December} 

\title{First measurement of symmetric cumulants of hexagonal flow harmonics in Pb--Pb collisions at $\mathbf{\sqrt{\textit{s}_{\mathrm{\mathbf{NN}}}}} \mathbf{\;= 5.02}$ TeV}

\ShortTitle{SC measurements of hexagonal flow harmonics in Pb--Pb collisions}   

\Collaboration{ALICE Collaboration\thanks{See Appendix~\ref{app:collab} for the list of collaboration members}}
\ShortAuthor{ALICE Collaboration} 

\begin{abstract}
Correlations between event-by-event fluctuations of anisotropic flow harmonics are measured in \mbox{Pb--Pb} collisions at a center-of-mass energy per nucleon pair of 5.02~TeV, as recorded by the ALICE detector at the LHC. This study presents correlations up to the hexagonal flow harmonic, $v_6$, which was measured for the first time. The magnitudes of these higher-order correlations are found to vary as a function of collision centrality and harmonic order. These measurements are compared to viscous hydrodynamic model calculations with EKRT initial conditions and to the iEBE-VISHNU model with {T\raisebox{-.5ex}{R}ENTo} initial conditions. The observed discrepancies between the data and the model calculations vary depending on the harmonic combinations. Due to the sensitivity of model parameters estimated with Bayesian analyses to these higher-order observables, the results presented in this work provide new and independent constraints on the initial conditions and transport properties in theoretical models used to describe the system created in heavy-ion collisions.

\end{abstract}
\end{titlepage}

\setcounter{page}{2} 


\section{Introduction}
\label{sec:intro}
The study of ultrarelativistic heavy-ion collisions aims to investigate the properties of the strongly interacting matter characterized by high energy densities and temperatures, known as quark--gluon plasma (QGP)~\cite{Braun-Munzinger:2015hba, Busza:2018rrf}.
These extreme conditions, needed for the production of QGP, can be achieved at the Relativistic Heavy Ion Collider (RHIC) at Brookhaven National Laboratory and at the Large Hadron Collider (LHC) at CERN.
Comparisons between experimental data and state-of-the-art model calculations have shown that the produced QGP is the most perfect fluid observed in nature so far, due to the small value of its shear viscosity over entropy density, $\eta/s$~\cite{Kovtun:2004de, ALICE:2022wpn}.
In recent years, one of the main focuses in heavy-ion collision studies has been determining the properties of the QGP using Bayesian analyses, which are designed to constrain parameters of the theoretical models via a comparison with different measured quantities~\cite{Bernhard:2016tnd, Bernhard:2019bmu, Auvinen:2020mpc, Nijs:2020ors, Nijs:2020roc, JETSCAPE:2020mzn, Parkkila:2021tqq, Parkkila:2021yha}.

One important probe of the QGP properties is the collective anisotropic flow, which translates the initial-state anisotropies in coordinate space into final-state anisotropies in the momentum distributions of produced particles~\cite{Ollitrault:1992bk}. Anisotropic flow is quantified by the flow amplitudes $v_n$ and the symmetry plane angles $\psi_n$ using a Fourier decomposition of the azimuthal distribution $f(\varphi)$ of the final-state particles in the plane transverse to the beam direction~\cite{Voloshin:1994mz},
\begin{equation}
f(\varphi)=\frac{1}{2\pi}\left[1+2\sum_{n=1}^{\infty}v_n\cos[n(\varphi-\psi_n)]\right]\,.
\label{eq:FourierSeries}
\end{equation}

Previous analyses have demonstrated that anisotropic flow is particularly sensitive to $\eta/s$ of the QGP~\cite{Voloshin:2008dg}. While only the final-state particles can be measured experimentally, it is possible to relate the observed flow coefficients to the initial-state spatial eccentricities defined as~\cite{Alver:2010gr, Gardim:2011xv}
\begin{equation}
    \varepsilon_n {\rm e}^{in\Phi_n} = -\frac{\{r^n {\rm e}^{in\varphi}\}}{\{r^n\}},\ n \geq 2.
\end{equation}
In Eq.~(2), the curly braces indicate an average defined by $\{\ldots\} = \int r {\rm d}r{\rm d}\varphi \varepsilon(r,\varphi)$ with $(r, \varphi)$ being the polar coordinates in the transverse plane, $\epsilon(r,\varphi)$ the initial energy density, and $\Phi_{n}$ represents the participant plane angle (see Refs.~\cite{Teaney:2010vd,Niemi:2015qia}).
It has been shown~\cite{Gardim:2011xv, Niemi:2012aj, Teaney:2013dta, Gardim:2014tya, Noronha-Hostler:2015dbi,Alver:2010dn,Lacey:2013eia} that the second- and third-order flow harmonics, $v_2$ and $v_3$, have linear contributions as well as non-linear dependencies from lower-order eccentricities~\cite{Qin:2010pf,Qiu:2011iv,Niemi:2012aj,Yan:2015jma}. 
More details on the expressions of these higher-order flow harmonics can be found in Refs.~\cite{ALICE:2020sup, ALICE:2023wdn}.

Experimental measurements~\cite{ALICE:2020sup, ALICE:2021adw, ALICE:2023wdn, ALICE:2023lwx} and theoretical calculations~\cite{Parkkila:2021tqq, Parkkila:2021yha} have demonstrated that observables related to the correlations between different flow harmonics are sensitive to the non-linear response, and in turn to the properties of the QGP~\cite{Parkkila:2021yha}. Only for small eccentricities the harmonics $v_n$ respond linearly to the eccentricities $\varepsilon_n$ of the same order, $v_n \propto \varepsilon_n$ (linear response), while for large eccentricities the anisotropies in momentum and coordinate space are interrelated via a matrix equation, which couples a set of anisotropic flow harmonics $\{v_n\}$ on one side, with the set of eccentricities $\{\varepsilon_n\}$ on the other (non-linear response). Later studies quantified the linear and non-linear contributions to $v_n$ and showed that the non-linear part becomes dominant in more peripheral collisions~\cite{ALICE:2020sup}. Recent Bayesian studies~\cite{Parkkila:2021tqq, Parkkila:2021yha} have measured the sensitivities to different observables used to constrain the model parameters, and concluded that the higher-order correlations are more sensitive to the medium properties than the ones used previously.

Correlations between different flow harmonics have been previously measured for harmonics $n$ ranging from $n=2$ to $5$~\cite{ALICE:2017kwu,Jia:2014jca,ATLAS:2015qwl}. This article extends the analysis of these correlations in Pb--Pb collisions at a center-of-mass energy per nucleon pair $\sqrt{{\textit s}_{\rm NN}}=5.02$~TeV up to the sixth order for the first time, building upon previous studies published in Refs.~\cite{ATLAS:2015qwl, ALICE:2016kpq, ALICE:2017kwu, ALICE:2021adw}.
The inclusion of the hexagonal flow harmonic $v_6$ is particularly interesting because of the different scaling of its non-linear response with the eccentricities $\epsilon_2$ and $\epsilon_3$ in the initial state (cubic $v_{6} \sim \varepsilon_2^3$ vs quadratic $v_{6} \sim \varepsilon_3^2$, respectively)~\cite{Yan:2015jma}. 

The article is organized as follows.
Section~\ref{sec:exp} introduces the experimental observables. The data analysis and systematic uncertainty evaluation is described in Sec.~\ref{sec:data} and the results are shown in Sec.~\ref{sec:results}. Finally, the main findings are summarized in Sec.~\ref{sec:summary}.

\section{Experimental observables}
\label{sec:exp}

While individual flow amplitudes and their event-by-event fluctuations provide valuable insight into the initial conditions, exploring correlations between different flow amplitudes can yield further independent constraints. Previous studies on these correlated fluctuations have led to the development of new observables~\cite{Niemi:2012aj, Bilandzic:2013kga, ATLAS:2015qwl, Qian:2016pau}.
For instance, the Symmetric Cumulants (SC) introduced by the \mbox{ALICE} Collaboration~\cite{Bilandzic:2013kga,ALICE:2016kpq,ALICE:2017kwu,Mordasini:2019hut,ALICE:2021klf} are direct multivariate cumulants of flow amplitudes, and each higher-order SC observable provides information that the lower-order ones cannot access. These observables are not dependent on the symmetry planes $\psi_n$ and are robust against systematic biases resulting from nonflow correlations (i.e.~correlations typically involving only a few particles, such as those induced by particle decays or jet fragmentation)~\cite{ALICE:2016kpq}. As reported in Refs.~\cite{Mordasini:2019hut,Bilandzic:2021rgb}, SC observables satisfy all fundamental mathematical and statistical properties of cumulants for any number and choice of flow harmonics.
Moreover, they are more sensitive to the temperature dependence of $\eta/s$ than individual flow amplitudes which primarily reflect the average values $\langle\eta/s\rangle$~\cite{Niemi:2015qia, ALICE:2016kpq}. In addition, it was demonstrated that these observables have the potential to disentangle contributions from initial conditions and medium properties, making it possible to directly constrain different stages in the evolution of heavy-ion collisions~\cite{Niemi:2015qia, ALICE:2016kpq}.
A recent state-of-the-art Bayesian analysis~\cite{Parkkila:2021yha} has quantified the sensitivity of the model parameters to all the observables included in the Bayesian estimation. This analysis is based on the {T\raisebox{-.5ex}{R}ENTo}+iEBE-VISHNU model~\cite{Moreland:2014oya}, which will be discussed in Sec.~\ref{sec:results}. The model is characterized by a total of 16 parameters, with key physics features embedded in the initial conditions, the temperature-dependent specific shear and bulk viscosity ($\eta/s(T)$ and $\zeta/s(T)$), free-streaming time ($\tau_{\rm fs}$), and switching temperature ($T_{\rm switch}$).
This study has also shown that the inclusion of the SC in the set of input observables has made it possible to reduce the uncertainties associated with the extracted medium properties.

Robust estimators for SC observables can be constructed experimentally using standard multiparticle azimuthal correlations~\cite{Bilandzic:2013kga, Mordasini:2019hut}.
In the case of two-harmonic SC, their definition is given by~\cite{Bilandzic:2013kga, ALICE:2016kpq, ALICE:2017kwu}
\begin{eqnarray}
\mathrm{SC}(m,n) \equiv \left<v_{m}^2v_{n}^2\right>-\left<v_{m}^2\right>\left<v_{n}^2\right>\ &=& \left<\left<\cos(m\varphi_1\!+\!n\varphi_2\!-\!m\varphi_3-\!n\varphi_4)\right>\right>\nonumber\\
&&{}-\left<\left<\cos[m(\varphi_1\!-\!\varphi_2)]\right>\right>\left<\left<\cos[n(\varphi_1\!-\!\varphi_2)]\right>\right> , 
\label{eq:4pCumulant}
\end{eqnarray}
with the condition $m \neq n$ for two positive integers $m$ and $n$. The double angular brackets indicate that the averaging is done in two separate steps. In the first step, all distinct particle
quadruplets in each event are formed and used to obtain single-event averages $\left< \cdots \right>$. In the second step, these single-event averages are weighted with ‘number of combinations’ weight to obtain the final all-event averages $\left<\left< \cdots \right>\right>$ (for further details, see Sec.~IV~C in Ref.~\cite{Bilandzic:2013kga}). It is crucial to define SC in terms of flow amplitudes $v_n$, to apply the cumulant expansion directly on $v_n$, and to use multiparticle azimuthal correlations only as estimators for each term in the resulting expression. This approach ensures the preservation of all the mathematical and statistical properties of the cumulants~\cite{Mordasini:2019hut,Bilandzic:2021rgb}. 
The resulting SC can be normalized by the product $\left<v_{m}^2\right>\left<v_{n}^2\right>$ using the following definition~\cite{ALICE:2016kpq, Giacalone:2016afq},
\begin{equation}
\mathrm{NSC}(m,n) \equiv \frac{\mathrm{SC}(m,n)}{\left<v_{m}^2\right>\left<v_{n}^2\right>}\,.
\label{eq:nsc}
\end{equation}
Normalized symmetric cumulants (NSC) allow for direct comparison of these observables in both momentum space (using $v_n$) and coordinate space (using $\epsilon_n$). This is due to the fact that the constant of proportionality which quantifies linear response in the relation $v_n \propto \varepsilon_n$ cancels exactly only in the NSC observables, and therefore a comparison of correlations in the initial coordinate and final momentum space can be performed at the same scale. An additional advantage of the NSC observables stems from the fact that any dependence of individual flow amplitudes on kinematic variables (e.g. transverse momentum $p_{\rm T}$) is suppressed, and it can be probed directly how the non-trivial patterns of correlations of flow harmonics change as a function of kinematic variables~\cite{ALICE:2017kwu}.

\begin{figure}[th!]
	\centering
            \includegraphics[width=1.0\textwidth]{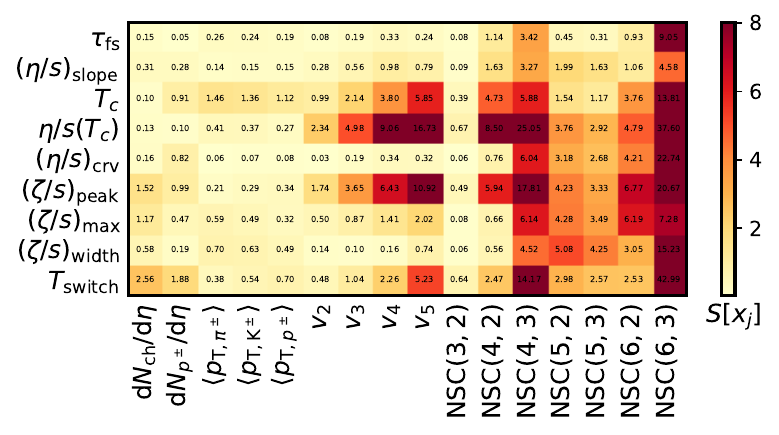}
	       \caption{Sensitivity of the model parameters in Bayesian analysis to the flow observables, shown as a color map. Light yellow shades represent low or no sensitivity, whereas orange and red colors represent moderate or strong sensitivities to the corresponding model parameter variation, respectively. The sensitivity analysis is based on the {T\raisebox{-.5ex}{R}ENTo}+iEBE-VISHNU model~\cite{Moreland:2014oya}. Several key parameters are displayed, including $\eta/s(T)$, $\zeta/s(T)$, $\tau_{\rm fs}$, and $T_{\rm switch}$. More details can be found in Ref.~\cite{Parkkila:2021yha}.}
  
 	      \label{fig:sensitivity}
\end{figure}

The sensitivity of model parameters to higher-order harmonic NSC was quantified using the method from Ref.~\cite{Parkkila:2021yha}, with results presented in Fig.~\ref{fig:sensitivity}. The sensitivity $S[x_j]$ of an observable $\hat{O}$ to parameter $x_j$ is defined as $|\hat{O}(\vec{x}')-\hat{O}(\vec{x})|/\delta \hat{O}(\vec{x})$. This measures how much an observable changes when a parameter is slightly changed by a constant value $\delta=0.1$. Here, $\hat{O}(\vec{x})$ represents the observable value at parameter point $\vec{x}=(x_1,\ldots,x_p)$, while $\vec{x}'$ denotes a point with a small change $\delta$ in $x_j$. The results are averaged across the centrality range of 5–30\%, where centrality is defined in terms of percentiles of the total hadronic cross section. This approach enables a quantitative assessment of which parameters most significantly impact the model's predictions. The analysis reveals that the model's transport properties are not highly sensitive to the number of charged particles or average transverse momentum $\langle p_\mathrm{T}\rangle$. However, the temperature-dependent specific shear viscosity, $\eta/s(T)$, is sensitive to $v_n$ values.
Notably, NSC exhibit even greater sensitivity to a wider range of parameters. This enhanced sensitivity suggests that NSC could be valuable to better constrain the parameters related to the transport properties of the medium.

\section{Data analysis}
\label{sec:data}
\subsection{Event and track selection}
\label{sec:selection}
This analysis utilizes data from Pb--Pb collisions at $\sqrt{s_{\rm NN}}= 5.02$~TeV recorded by the ALICE detector in 2015 and 2018. The ALICE detector includes several subdetectors immersed in a 0.5~T solenoidal field. The Inner Tracking System (ITS)~\cite{ALICE:1999cls, ALICE:2010tia} is used for track reconstruction. Positioned closest to the beam vacuum tube, the ITS consists of six silicon layers with three types of detector technologies. The two innermost layers, Silicon Pixel Detectors (SPD), provide high spatial granularity which is ideal for reconstructing primary and secondary vertices. Surrounding the ITS is the Time Projection Chamber (TPC)~\cite{Alme:2010ke}, a gas-filled cylindrical tracking detector that provides up to 159 reconstruction points for charged tracks traversing its full radial extent. It is used for reconstructing charged-particle tracks and for particle identification. The detailed descriptions of the various detectors and their performance are given in Refs.~\cite{ALICE:2008ngc, ALICE:2014sbx}.

Triggering and centrality determination is carried out using two scintillator arrays, V0A and V0C~\cite{ALICE:2008ngc, ALICE:2013axi}. The centrality determined using these V0 detectors is referred to as the V0 estimator. All three detectors (TPC, ITS, and the V0 arrays) cover the full azimuth. They have pseudorapidity ranges within $|\eta| < 0.9$ for the TPC and ITS, and $2.8 < \eta < 5.1$ and $-3.7 < \eta < -1.7$ for V0A and V0C, respectively.

Minimum bias (MB) events are triggered by a coincident signal in both the V0A and V0C. Only MB Pb--Pb events with a reconstructed primary vertex within $\pm 8.0$~cm from the nominal interaction point along the beam direction are selected. To remove background events such as beam--gas collisions and pile-up, information from the V0 detector and the SPD is utilized as done in Ref.~\cite{ALICE:2024fus}. After applying all event selection criteria, 212 million events remain within the 0–60\% centrality range.

This analysis involves tracks reconstructed using combined information from the ITS and TPC within a transverse momentum interval of $0.2<p_{\rm T}<5.0$~GeV/$c$ and a pseudorapidity range of $|\eta| < 0.8$. To avoid contributions from secondary particles, only tracks with a specified distance of closest approach (DCA) to the primary vertex are accepted. Furthermore, the reconstructed tracks are required to have a minimum of 70 TPC space points and a minimum of 2 hits in the ITS. All kink topology tracks are rejected. The selection criteria employed in this analysis align closely with those outlined in Refs.~\cite{ALICE:2022zks,ALICE:2023lwx}. After these track selections, an extra criterion is enforced to discard any remaining events with fewer than 10 reconstructed tracks as this is the smallest number of tracks necessary for calculating all relevant SC$(m,n)$ observables, as done in Ref.~\cite{ALICE:2023lwx}.

Corrections for the non-uniform reconstruction efficiency (NUE) and the non-uniform acceptance (NUA) are applied as a function of transverse momentum and as a function of azimuthal angle, respectively, following previous studies~\cite{Bilandzic:2013kga, ALICE:2020sup, ALICE:2021klf}. The NUA correction is data-driven, while the NUE correction factor is calculated with a Monte Carlo simulation using the HIJING~\cite{Gyulassy:1994ew} event generator and GEANT3~\cite{Brun:1994aa} transport software, accounting for the track reconstruction efficiency and contamination from secondary particles.

In order to suppress the nonflow contribution resulting from the two-particle correlations in the denominator of the NSC in Eq.~(\ref{eq:nsc}), a pseudorapidity gap of $|\Delta\eta|> 1.0$ is used. 
For the two two-particle correlations which appear in the definition of SC$(m,n)$ in Eq.~(\ref{eq:4pCumulant}), the pseudorapidity gap is not needed, since nonflow is suppressed by construction of this observable. This was demonstrated by HIJING model~\cite{Gyulassy:1994ew} simulations and the like-sign technique in Ref.~\cite{ALICE:2016kpq}.

\subsection{Systematic uncertainties}
\label{sec:syst}
The systematic uncertainties are estimated by varying the event and track selection criteria with respect to the default selections, previously summarized, taking into account the correlations between their statistical uncertainties as done in Ref.~\cite{ALICE:2023lwx}. 
Each selection criterion variation is described below.

The effect of the centrality determination is estimated by changing the default V0 estimator to the SPD. The selection on the longitudinal position of the primary vertex is varied from $\pm$8 cm to $\pm$7 and $\pm$9 cm. About 20\% of the data collected in 2015 and over 25\% of the data collected in 2018 were affected by out-of-bunch pile-up collisions. The effects of pile-up collisions are studied using correlations between the number of tracks measured in the TPC and the number of tracks reconstructed with the ITS to reduce contamination from events occurring in different bunch crossings~\cite{Arslandok:2022dyb}.
The impact of the two configurations of the magnetic field polarity in the solenoid magnet of ALICE is investigated by performing the analysis on the data sets taken for each orientation separately.
To test the track-quality selections, the minimum number of space points in the TPC required for the track reconstruction is changed from 70 to 80 and 65. The $\chi^2$ value per space point from the track fit is reduced from 2.5 to 2.3. Finally, the DCA of the extrapolated track to the primary vertex position is tightened from 2 cm to 1 cm along the beam direction, while in the transverse plane a transverse-momentum dependent DCA selection was applied to account for the $\pt$ dependence of the DCA resolution (using the expression $0.0208+0.04/p_{\rm T}^{1.1}$~cm, with $p_{\rm T}$ expressed in units of GeV/$c$). 

The significance of the difference for each variation is determined using the Barlow test~\cite{Barlow:2002yb}. If the statistical significance of the performed systematic variation is found to be greater than a certain value, that systematic variation is classified as a statistically relevant deviation, and is added to the evaluation of total systematic uncertainty. A Barlow criterion of 2.0 is applied for all observables except NSC$(6,2)$ and NSC$(6,3)$. For NSC$(6,2)$, a relaxed criterion of 1.0 is used to assign the systematic uncertainty. In the case of NSC$(6,3)$, due to large statistical uncertainties, the Barlow test is not applied. Instead, all trials contribute to the systematic uncertainty, except for the variation in magnetic field polarity due to large statistical uncertainties of a given dataset with different polarity.
It is important to note that the 2015 and 2018 datasets were collected under slightly different detector conditions~\cite{Arslandok:2022dyb}, which particularly affected the 0--5\% and 5--10\% centrality percentiles, where track multiplicity is the highest. Therefore, the two data sets were analyzed separately, and the observed differences were assigned as an additional source of systematic uncertainty, which is the dominant source of systematics for 0–10\% centrality percentiles.
The variations from each systematic source are added in quadrature to obtain the total systematic uncertainties.

\section{Results}
\label{sec:results}

\begin{figure}[th!]
	\centering
            \includegraphics[width=1.0\textwidth]{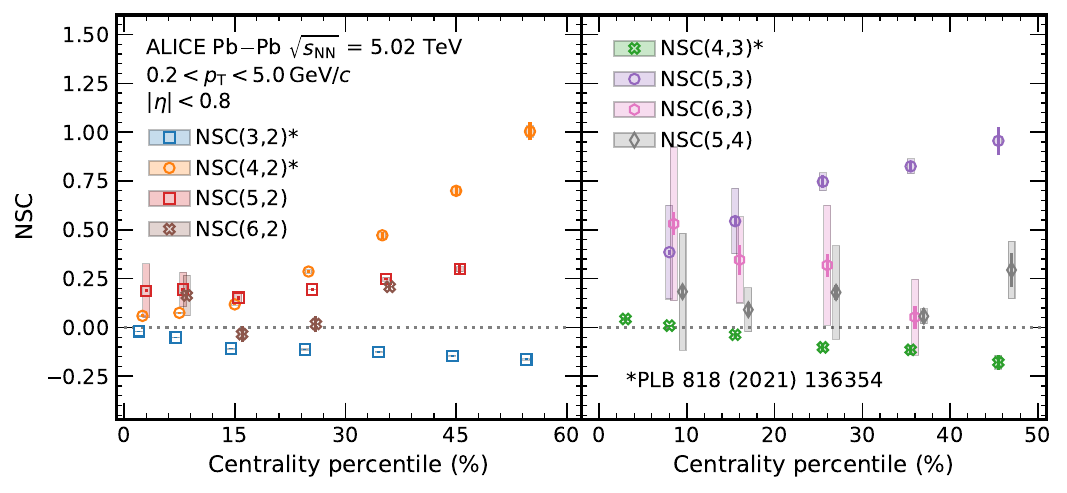}
	       \caption{The centrality dependence of the NSC observables. The previously published lower-order harmonics~\cite{ALICE:2021adw} are marked with asterisk (*). The statistical and total systematic uncertainties are shown with vertical lines and boxes, respectively.}
  
 	      \label{fig:NSCresults}
\end{figure}
The results of the higher-order harmonic normalized symmetric cumulants are shown in Fig.~\ref{fig:NSCresults} together with the lower-order NSC from Ref.~\cite{ALICE:2021adw} marked with an asterisk (*). The new measurements include the fifth and sixth harmonic amplitudes.
All observables are positive except for NSC(3,2) and for NSC(4,3), in non-central collisions. In general, the sign of two-harmonic (N)SC observables has a non-trivial physics interpretation, and can be understood as follows. A positive (N)SC indicates that measuring $v_m$ larger than $\langle v_m\rangle$ in an event will increase the probability of measuring $v_n$ larger than $\langle v_n\rangle$ in that event, i.e.~event-by-event fluctuations of $v_m$ and $v_n$ are correlated. On the other hand, a negative sign can be interpreted as anti-correlation between the event-by-event fluctuations of $v_m$ and $v_n$ amplitudes, meaning that measuring $v_m$ larger than $\langle v_m\rangle$ will decrease the probability of measuring $v_n$ larger than $\langle v_n\rangle$ in the same event~\cite{ALICE:2016kpq}. The physical interpretation of the sign of higher-order SC observables involving more than two harmonics is more challenging, and can be found in Ref.~\cite{ALICE:2021klf}. 

In Fig.~\ref{fig:NSCresults}, a strong centrality dependence is observed for NSC(5,3), along with a possible hint for a slightly decreasing trend for NSC(6,3) toward peripheral collisions, while no significant centrality dependence is seen for NSC(5,2), NSC(5,4) and NSC(6,2). The different trends of the different NSC observables from most central to semicentral collisions are expected due to the fact that the physical origin of the flow fluctuations is different in these two regimes. While the main driving force in the most central collisions are fluctuations of participating nucleons, in semicentral collisions fluctuations are of geometric origin due to the leading-order ellipsoidal shape. The centrality dependence of NSC(6,2) and NSC(6,3) is qualitatively different, despite large uncertainties for NSC(6,3). This difference is not surprising given the distinct scaling of the non-linear response contribution in these two cases ($v_6 \sim v_2^3$ and $v_6 \sim v_3^2$, respectively)~\cite{Yan:2015jma}. Such different centrality dependence demonstrates that for different combinations of flow harmonics, NSC observables extract new and independent information about heavy-ion collisions.

\begin{figure}[tbh!]
	\centering
\includegraphics[width=0.99\textwidth]{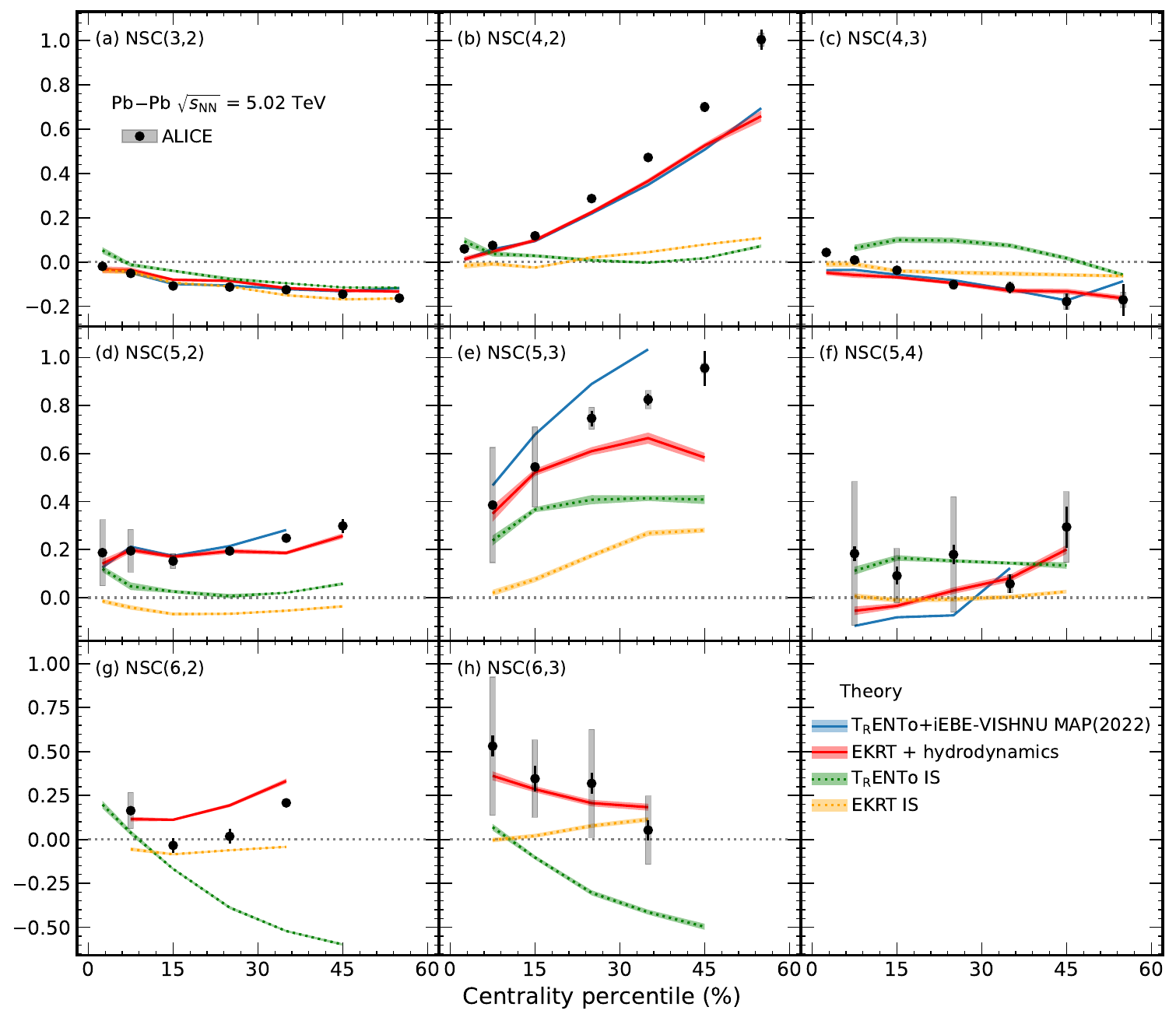}
	\caption{The centrality dependence of NSC($m$,$n$) in $\PbPb$ collisions at $\snn=5.02$~TeV. Results for each observable are compared with the final-state predictions from the event-by-event EKRT~+~viscous hydrodynamic calculations~\cite{Niemi:2015qia} and {T\raisebox{-.5ex}{R}ENTo}+iEBE-VISHNU MAP (2022), as well as with the initial-state calculations from EKRT and {T\raisebox{-.5ex}{R}ENTo}. Panels (a) to (c) present data from Ref.~\cite{ALICE:2021adw}. The statistical and total systematic uncertainties of the data are shown with vertical lines and boxes, respectively. The model results are shown as colored bands with the width of the band denoting the statistical uncertainties.}
 	\label{fig:NSC_EKRT}
\end{figure}

A systematic comparison of the centrality dependence of the NSC($m$,$n$) to initial- and final-state models has been performed and it is shown in Fig.~\ref{fig:NSC_EKRT}. For the comparison to the EKRT and {T\raisebox{-.5ex}{R}ENTo} initial-state models, the observable is calculated using the initial state eccentricities. The data are also compared with the EKRT~+~viscous hydrodynamics model~\cite{Niemi:2015qia} and the TRENTo+iEBE-VISHNU model ~\cite{Moreland:2014oya}. 

In the event-by-event EKRT~+~viscous hydrodynamic calculations, the initial energy density profiles are calculated using a next-to-leading order perturbative quantum chromodynamics approach implemented with gluon saturation model~\cite{Paatelainen:2012at,Paatelainen:2013eea}. The subsequent space--time evolution is described by relativistic dissipative fluid dynamics with different parameterizations for the temperature dependence of the shear viscosity to entropy density ratio, $\eta/s(T)$. This model gives a good description of the charged hadron multiplicity and the low-$p_{\rm T}$ region of the charged hadron spectra at RHIC and the LHC (see Figs.~11--13 in Ref.~\cite{Niemi:2015qia}). Each $\eta/s(T)$ parameterization is tuned to reproduce the measured $v_2$ from central to semiperipheral collisions (see Fig.~10 in Ref.~\cite{Hirvonen:2022xfv}), while keeping the average $\langle\eta/s(T)\rangle$ the same for all parameterizations.

The {T\raisebox{-.5ex}{R}ENTo}+iEBE-VISHNU model uses {T\raisebox{-.5ex}{R}ENTo}~\cite{Moreland:2014oya} to simulate the initial conditions, which are then connected with a free streaming phase transitioning into a 2+1 dimensional causal hydrodynamic model known as VISH2+1~\cite{Shen:2014vra}. The hydrodynamic evolution within VISH2+1 accounts for the expansion and cooling of the QGP, leading up to hadronization. After hadronization, the evolution continues using a hadronic cascade model (UrQMD)~\cite{Bass:1998ca,Bleicher:1999xi}, which simulates the interactions and decays of the produced hadrons, ensuring a realistic description of the final-state particles. The model calculation uses the best-fit parameterization for transport coefficients selected based on maximum a posteriori (MAP) for Pb--Pb collisions at $\sqrt{s_{\text{NN}}}=5.02$~TeV. The MAP values are based on Ref.~\cite{Parkkila:2021yha} and labeled MAP(2022) in Fig.~\ref{fig:NSC_EKRT}.

Figure~\ref{fig:NSC_EKRT} presents all the measured NSC's, where panels (a) to (c) show the data from Ref.~\cite{ALICE:2021adw}.
A good agreement between the final-state models, the EKRT~+~viscous hydrodynamics model~\cite{Hirvonen:2022xfv} and {T\raisebox{-.5ex}{R}ENTo}+iEBE-VISHNU, and data can be seen for NSC(3,2), NSC(4,3), NSC(5,2), and NSC(5,4).
The comparison between initial and final states in the models reveals that, with the exceptions of NSC(3,2) and NSC(5,4), all other observables demonstrate a pronounced non-linear hydrodynamic response. This response significantly outweighs the influence of initial-state eccentricities. 
Most observables show an increasing correlation in peripheral collisions, with two exceptions: NSC(6,3) and NSC(5,4). NSC(5,4) demonstrates minimal centrality dependence within the uncertainties. NSC(6,3), however, exhibits a distinct centrality dependence. This suggests a negative contribution from NSC(3,2) coupled with an increasing correlation towards peripheral collisions, which results in a decreasing trend in peripheral collisions. In summary, the discrepancies between experimental data and model calculations vary across different harmonic combinations. The two models for the initial state, EKRT and {T\raisebox{-.5ex}{R}ENTo}, show distinct initial-state correlations for most observables. Additionally, the hydrodynamic response to these initial states differs between the models, leading to distinct predictions for the final-state observables. These variations offer valuable opportunities to constrain both initial conditions and final-state effects in heavy-ion collision models.

\section{Summary}
\label{sec:summary}
In conclusion, the first measurements of higher-order harmonic NSC(6,2) and NSC(6,3) in Pb--Pb collisions at $\sqrt{s_{\mathrm{NN}}}=5.02$~TeV are reported.
 These observables are crucial for further constraining theoretical models, as they provide new and independent information on the evolution of a heavy-ion collision. It was also demonstrated that the model parameterization in Bayesian analysis is sensitive to higher-order harmonic NSC($m,n$) and can therefore be used to decrease the uncertainty on the transport properties of the QGP.
These newly measured observables and the lower-order harmonic NSC($m,n$) are compared with hydrodynamic calculations, where, generally, the agreement with data is worse for higher-order NSC($m,n$). It is also observed that both NSC(6,2) and NSC(6,3) are positive, and that NSC(6,3) is the only observable with hint of a decreasing correlation with increasing centrality.
Discrepancies between data and model calculations vary across harmonic combinations. The EKRT and iEBE-VISHNU with {T\raisebox{-.5ex}{R}ENTo} models show distinct initial state correlations and hydrodynamic responses. These variations help constrain initial conditions and final-state effects in heavy-ion collision models. Moreover, the sensitivity of Bayesian model parameterization to these higher-order observables offers new constraints on initial conditions and transport properties in theoretical models of heavy-ion collisions.


\newenvironment{acknowledgement}{\relax}{\relax}
\begin{acknowledgement}
\section*{Acknowledgements}
The ALICE Collaboration extends its gratitude to Henry Hirvonen for providing the latest predictions from the state-of-the-art hydrodynamic model, event-by-event EKRT+hydrodynamics. We also thank the Jyväskylä group for supplying the sensitivity map figure.

The ALICE Collaboration would like to thank all its engineers and technicians for their invaluable contributions to the construction of the experiment and the CERN accelerator teams for the outstanding performance of the LHC complex.
The ALICE Collaboration gratefully acknowledges the resources and support provided by all Grid centres and the Worldwide LHC Computing Grid (WLCG) collaboration.
The ALICE Collaboration acknowledges the following funding agencies for their support in building and running the ALICE detector:
A. I. Alikhanyan National Science Laboratory (Yerevan Physics Institute) Foundation (ANSL), State Committee of Science and World Federation of Scientists (WFS), Armenia;
Austrian Academy of Sciences, Austrian Science Fund (FWF): [M 2467-N36] and Nationalstiftung f\"{u}r Forschung, Technologie und Entwicklung, Austria;
Ministry of Communications and High Technologies, National Nuclear Research Center, Azerbaijan;
Conselho Nacional de Desenvolvimento Cient\'{\i}fico e Tecnol\'{o}gico (CNPq), Financiadora de Estudos e Projetos (Finep), Funda\c{c}\~{a}o de Amparo \`{a} Pesquisa do Estado de S\~{a}o Paulo (FAPESP) and Universidade Federal do Rio Grande do Sul (UFRGS), Brazil;
Bulgarian Ministry of Education and Science, within the National Roadmap for Research Infrastructures 2020-2027 (object CERN), Bulgaria;
Ministry of Education of China (MOEC) , Ministry of Science \& Technology of China (MSTC) and National Natural Science Foundation of China (NSFC), China;
Ministry of Science and Education and Croatian Science Foundation, Croatia;
Centro de Aplicaciones Tecnol\'{o}gicas y Desarrollo Nuclear (CEADEN), Cubaenerg\'{\i}a, Cuba;
Ministry of Education, Youth and Sports of the Czech Republic, Czech Republic;
The Danish Council for Independent Research | Natural Sciences, the VILLUM FONDEN and Danish National Research Foundation (DNRF), Denmark;
Helsinki Institute of Physics (HIP), Finland;
Commissariat \`{a} l'Energie Atomique (CEA) and Institut National de Physique Nucl\'{e}aire et de Physique des Particules (IN2P3) and Centre National de la Recherche Scientifique (CNRS), France;
Bundesministerium f\"{u}r Bildung und Forschung (BMBF) and GSI Helmholtzzentrum f\"{u}r Schwerionenforschung GmbH, Germany;
General Secretariat for Research and Technology, Ministry of Education, Research and Religions, Greece;
National Research, Development and Innovation Office, Hungary;
Department of Atomic Energy Government of India (DAE), Department of Science and Technology, Government of India (DST), University Grants Commission, Government of India (UGC) and Council of Scientific and Industrial Research (CSIR), India;
National Research and Innovation Agency - BRIN, Indonesia;
Istituto Nazionale di Fisica Nucleare (INFN), Italy;
Japanese Ministry of Education, Culture, Sports, Science and Technology (MEXT) and Japan Society for the Promotion of Science (JSPS) KAKENHI, Japan;
Consejo Nacional de Ciencia (CONACYT) y Tecnolog\'{i}a, through Fondo de Cooperaci\'{o}n Internacional en Ciencia y Tecnolog\'{i}a (FONCICYT) and Direcci\'{o}n General de Asuntos del Personal Academico (DGAPA), Mexico;
Nederlandse Organisatie voor Wetenschappelijk Onderzoek (NWO), Netherlands;
The Research Council of Norway, Norway;
Pontificia Universidad Cat\'{o}lica del Per\'{u}, Peru;
Ministry of Science and Higher Education, National Science Centre and WUT ID-UB, Poland;
Korea Institute of Science and Technology Information and National Research Foundation of Korea (NRF), Republic of Korea;
Ministry of Education and Scientific Research, Institute of Atomic Physics, Ministry of Research and Innovation and Institute of Atomic Physics and Universitatea Nationala de Stiinta si Tehnologie Politehnica Bucuresti, Romania;
Ministry of Education, Science, Research and Sport of the Slovak Republic, Slovakia;
National Research Foundation of South Africa, South Africa;
Swedish Research Council (VR) and Knut \& Alice Wallenberg Foundation (KAW), Sweden;
European Organization for Nuclear Research, Switzerland;
Suranaree University of Technology (SUT), National Science and Technology Development Agency (NSTDA) and National Science, Research and Innovation Fund (NSRF via PMU-B B05F650021), Thailand;
Turkish Energy, Nuclear and Mineral Research Agency (TENMAK), Turkey;
National Academy of  Sciences of Ukraine, Ukraine;
Science and Technology Facilities Council (STFC), United Kingdom;
National Science Foundation of the United States of America (NSF) and United States Department of Energy, Office of Nuclear Physics (DOE NP), United States of America.
In addition, individual groups or members have received support from:
Czech Science Foundation (grant no. 23-07499S), Czech Republic;
FORTE project, reg.\ no.\ CZ.02.01.01/00/22\_008/0004632, Czech Republic, co-funded by the European Union, Czech Republic;
European Research Council (grant no. 950692), European Union;
Academy of Finland (Center of Excellence in Quark Matter) (grant nos. 346327, 346328), Finland;
Deutsche Forschungs Gemeinschaft (DFG, German Research Foundation) ``Neutrinos and Dark Matter in Astro- and Particle Physics'' (grant no. SFB 1258), Germany;
ICSC - National Research Center for High Performance Computing, Big Data and Quantum Computing and FAIR - Future Artificial Intelligence Research, funded by the NextGenerationEU program (Italy).

\end{acknowledgement}

\bibliographystyle{utphys}   
\bibliography{bibliography}

\newpage
\appendix

%
%

\section{The ALICE Collaboration}
\label{app:collab}
\begin{flushleft} 
\small

S.~Acharya\,\orcidlink{0000-0002-9213-5329}\,$^{\rm 50}$, 
A.~Agarwal$^{\rm 133}$, 
G.~Aglieri Rinella\,\orcidlink{0000-0002-9611-3696}\,$^{\rm 32}$, 
L.~Aglietta\,\orcidlink{0009-0003-0763-6802}\,$^{\rm 24}$, 
M.~Agnello\,\orcidlink{0000-0002-0760-5075}\,$^{\rm 29}$, 
N.~Agrawal\,\orcidlink{0000-0003-0348-9836}\,$^{\rm 25}$, 
Z.~Ahammed\,\orcidlink{0000-0001-5241-7412}\,$^{\rm 133}$, 
S.~Ahmad\,\orcidlink{0000-0003-0497-5705}\,$^{\rm 15}$, 
S.U.~Ahn\,\orcidlink{0000-0001-8847-489X}\,$^{\rm 71}$, 
I.~Ahuja\,\orcidlink{0000-0002-4417-1392}\,$^{\rm 36}$
% ...

\section*{Affiliation Notes}

$^{\rm I}$ Deceased\\
$^{\rm II}$ Also at: Max-Planck-Institut fur Physik, Munich, Germany\\
$^{\rm III}$ Also at: Italian National Agency for New Technologies, Energy and Sustainable Economic Development (ENEA), Bologna, Italy\\
$^{\rm IV}$ Also at: Dipartimento DET del Politecnico di Torino, Turin, Italy\\
$^{\rm V}$ Also at: Department of Applied Physics, Aligarh Muslim University, Aligarh, India\\
$^{\rm VI}$ Also at: Institute of Theoretical Physics, University of Wroclaw, Poland\\
$^{\rm VII}$ Also at: Facultad de Ciencias, Universidad Nacional Aut\'{o}noma de M\'{e}xico, Mexico City, Mexico\\

\section*{Collaboration Institutes}

$^{1}$ A.I. Alikhanyan National Science Laboratory (Yerevan Physics Institute) Foundation, Yerevan, Armenia\\
$^{2}$ AGH University of Krakow, Cracow, Poland\\
$^{3}$ Bogolyubov Institute for Theoretical Physics, National Academy of Sciences of Ukraine, Kiev, Ukraine\\
$^{4}$ Bose Institute, Department of Physics  and Centre for Astroparticle Physics and Space Science (CAPSS), Kolkata, India\\
$^{5}$ California Polytechnic State University, San Luis Obispo, California, United States\\
$^{6}$ Central China Normal University, Wuhan, China\\
$^{7}$ Centro de Aplicaciones Tecnol\'{o}gicas y Desarrollo Nuclear (CEADEN), Havana, Cuba\\
$^{8}$ Centro de Investigaci\'{o}n y de Estudios Avanzados (CINVESTAV), Mexico City and M\'{e}rida, Mexico\\
$^{9}$ Chicago State University, Chicago, Illinois, United States\\
$^{10}$ China Nuclear Data Center, China Institute of Atomic Energy, Beijing, China\\
$^{11}$ China University of Geosciences, Wuhan, China\\
$^{12}$ Chungbuk National University, Cheongju, Republic of Korea\\
$^{13}$ Comenius University Bratislava, Faculty of Mathematics, Physics and Informatics, Bratislava, Slovak Republic\\
$^{14}$ Creighton University, Omaha, Nebraska, United States\\
$^{15}$ Department of Physics, Aligarh Muslim University, Aligarh, India\\
$^{16}$ Department of Physics, Pusan National University, Pusan, Republic of Korea\\
$^{17}$ Department of Physics, Sejong University, Seoul, Republic of Korea\\
$^{18}$ Department of Physics, University of California, Berkeley, California, United States\\
$^{19}$ Department of Physics, University of Oslo, Oslo, Norway\\
$^{20}$ Department of Physics and Technology, University of Bergen, Bergen, Norway\\
$^{21}$ Dipartimento di Fisica, Universit\`{a} di Pavia, Pavia, Italy\\
$^{22}$ Dipartimento di Fisica dell'Universit\`{a} and Sezione INFN, Cagliari, Italy\\
$^{23}$ Dipartimento di Fisica dell'Universit\`{a} and Sezione INFN, Trieste, Italy\\
$^{24}$ Dipartimento di Fisica dell'Universit\`{a} and Sezione INFN, Turin, Italy\\
$^{25}$ Dipartimento di Fisica e Astronomia dell'Universit\`{a} and Sezione INFN, Bologna, Italy\\
$^{26}$ Dipartimento di Fisica e Astronomia dell'Universit\`{a} and Sezione INFN, Catania, Italy\\
$^{27}$ Dipartimento di Fisica e Astronomia dell'Universit\`{a} and Sezione INFN, Padova, Italy\\
$^{28}$ Dipartimento di Fisica `E.R.~Caianiello' dell'Universit\`{a} and Gruppo Collegato INFN, Salerno, Italy\\
$^{29}$ Dipartimento DISAT del Politecnico and Sezione INFN, Turin, Italy\\
$^{30}$ Dipartimento di Scienze MIFT, Universit\`{a} di Messina, Messina, Italy\\
$^{31}$ Dipartimento Interateneo di Fisica `M.~Merlin' and Sezione INFN, Bari, Italy\\
$^{32}$ European Organization for Nuclear Research (CERN), Geneva, Switzerland\\
$^{33}$ Faculty of Electrical Engineering, Mechanical Engineering and Naval Architecture, University of Split, Split, Croatia\\
$^{34}$ Faculty of Nuclear Sciences and Physical Engineering, Czech Technical University in Prague, Prague, Czech Republic\\
$^{35}$ Faculty of Physics, Sofia University, Sofia, Bulgaria\\
$^{36}$ Faculty of Science, P.J.~\v{S}af\'{a}rik University, Ko\v{s}ice, Slovak Republic\\
$^{37}$ Faculty of Technology, Environmental and Social Sciences, Bergen, Norway\\
$^{38}$ Frankfurt Institute for Advanced Studies, Johann Wolfgang Goethe-Universit\"{a}t Frankfurt, Frankfurt, Germany\\
$^{39}$ Fudan University, Shanghai, China\\
$^{40}$ Gangneung-Wonju National University, Gangneung, Republic of Korea\\
$^{41}$ Gauhati University, Department of Physics, Guwahati, India\\
$^{42}$ Helmholtz-Institut f\"{u}r Strahlen- und Kernphysik, Rheinische Friedrich-Wilhelms-Universit\"{a}t Bonn, Bonn, Germany\\
$^{43}$ Helsinki Institute of Physics (HIP), Helsinki, Finland\\
$^{44}$ High Energy Physics Group,  Universidad Aut\'{o}noma de Puebla, Puebla, Mexico\\
$^{45}$ Horia Hulubei National Institute of Physics and Nuclear Engineering, Bucharest, Romania\\
$^{46}$ HUN-REN Wigner Research Centre for Physics, Budapest, Hungary\\
$^{47}$ Indian Institute of Technology Bombay (IIT), Mumbai, India\\
$^{48}$ Indian Institute of Technology Indore, Indore, India\\
$^{49}$ INFN, Laboratori Nazionali di Frascati, Frascati, Italy\\
$^{50}$ INFN, Sezione di Bari, Bari, Italy\\
$^{51}$ INFN, Sezione di Bologna, Bologna, Italy\\
$^{52}$ INFN, Sezione di Cagliari, Cagliari, Italy\\
$^{53}$ INFN, Sezione di Catania, Catania, Italy\\
$^{54}$ INFN, Sezione di Padova, Padova, Italy\\
$^{55}$ INFN, Sezione di Pavia, Pavia, Italy\\
$^{56}$ INFN, Sezione di Torino, Turin, Italy\\
$^{57}$ INFN, Sezione di Trieste, Trieste, Italy\\
$^{58}$ Inha University, Incheon, Republic of Korea\\
$^{59}$ Institute for Gravitational and Subatomic Physics (GRASP), Utrecht University/Nikhef, Utrecht, Netherlands\\
$^{60}$ Institute of Experimental Physics, Slovak Academy of Sciences, Ko\v{s}ice, Slovak Republic\\
$^{61}$ Institute of Physics, Homi Bhabha National Institute, Bhubaneswar, India\\
$^{62}$ Institute of Physics of the Czech Academy of Sciences, Prague, Czech Republic\\
$^{63}$ Institute of Space Science (ISS), Bucharest, Romania\\
$^{64}$ Institut f\"{u}r Kernphysik, Johann Wolfgang Goethe-Universit\"{a}t Frankfurt, Frankfurt, Germany\\
$^{65}$ Instituto de Ciencias Nucleares, Universidad Nacional Aut\'{o}noma de M\'{e}xico, Mexico City, Mexico\\
$^{66}$ Instituto de F\'{i}sica, Universidade Federal do Rio Grande do Sul (UFRGS), Porto Alegre, Brazil\\
$^{67}$ Instituto de F\'{\i}sica, Universidad Nacional Aut\'{o}noma de M\'{e}xico, Mexico City, Mexico\\
$^{68}$ iThemba LABS, National Research Foundation, Somerset West, South Africa\\
$^{69}$ Jeonbuk National University, Jeonju, Republic of Korea\\
$^{70}$ Johann-Wolfgang-Goethe Universit\"{a}t Frankfurt Institut f\"{u}r Informatik, Fachbereich Informatik und Mathematik, Frankfurt, Germany\\
$^{71}$ Korea Institute of Science and Technology Information, Daejeon, Republic of Korea\\
$^{72}$ Laboratoire de Physique Subatomique et de Cosmologie, Universit\'{e} Grenoble-Alpes, CNRS-IN2P3, Grenoble, France\\
$^{73}$ Lawrence Berkeley National Laboratory, Berkeley, California, United States\\
$^{74}$ Lund University Department of Physics, Division of Particle Physics, Lund, Sweden\\
$^{75}$ Nagasaki Institute of Applied Science, Nagasaki, Japan\\
$^{76}$ Nara Women{'}s University (NWU), Nara, Japan\\
$^{77}$ National and Kapodistrian University of Athens, School of Science, Department of Physics , Athens, Greece\\
$^{78}$ National Centre for Nuclear Research, Warsaw, Poland\\
$^{79}$ National Institute of Science Education and Research, Homi Bhabha National Institute, Jatni, India\\
$^{80}$ National Nuclear Research Center, Baku, Azerbaijan\\
$^{81}$ National Research and Innovation Agency - BRIN, Jakarta, Indonesia\\
$^{82}$ Niels Bohr Institute, University of Copenhagen, Copenhagen, Denmark\\
$^{83}$ Nikhef, National institute for subatomic physics, Amsterdam, Netherlands\\
$^{84}$ Nuclear Physics Group, STFC Daresbury Laboratory, Daresbury, United Kingdom\\
$^{85}$ Nuclear Physics Institute of the Czech Academy of Sciences, Husinec-\v{R}e\v{z}, Czech Republic\\
$^{86}$ Oak Ridge National Laboratory, Oak Ridge, Tennessee, United States\\
$^{87}$ Ohio State University, Columbus, Ohio, United States\\
$^{88}$ Physics department, Faculty of science, University of Zagreb, Zagreb, Croatia\\
$^{89}$ Physics Department, Panjab University, Chandigarh, India\\
$^{90}$ Physics Department, University of Jammu, Jammu, India\\
$^{91}$ Physics Program and International Institute for Sustainability with Knotted Chiral Meta Matter (WPI-SKCM$^{2}$), Hiroshima University, Hiroshima, Japan\\
$^{92}$ Physikalisches Institut, Eberhard-Karls-Universit\"{a}t T\"{u}bingen, T\"{u}bingen, Germany\\
$^{93}$ Physikalisches Institut, Ruprecht-Karls-Universit\"{a}t Heidelberg, Heidelberg, Germany\\
$^{94}$ Physik Department, Technische Universit\"{a}t M\"{u}nchen, Munich, Germany\\
$^{95}$ Politecnico di Bari and Sezione INFN, Bari, Italy\\
$^{96}$ Research Division and ExtreMe Matter Institute EMMI, GSI Helmholtzzentrum f\"ur Schwerionenforschung GmbH, Darmstadt, Germany\\
$^{97}$ Saga University, Saga, Japan\\
$^{98}$ Saha Institute of Nuclear Physics, Homi Bhabha National Institute, Kolkata, India\\
$^{99}$ School of Physics and Astronomy, University of Birmingham, Birmingham, United Kingdom\\
$^{100}$ Secci\'{o}n F\'{\i}sica, Departamento de Ciencias, Pontificia Universidad Cat\'{o}lica del Per\'{u}, Lima, Peru\\
$^{101}$ Stefan Meyer Institut f\"{u}r Subatomare Physik (SMI), Vienna, Austria\\
$^{102}$ SUBATECH, IMT Atlantique, Nantes Universit\'{e}, CNRS-IN2P3, Nantes, France\\
$^{103}$ Sungkyunkwan University, Suwon City, Republic of Korea\\
$^{104}$ Suranaree University of Technology, Nakhon Ratchasima, Thailand\\
$^{105}$ Technical University of Ko\v{s}ice, Ko\v{s}ice, Slovak Republic\\
$^{106}$ The Henryk Niewodniczanski Institute of Nuclear Physics, Polish Academy of Sciences, Cracow, Poland\\
$^{107}$ The University of Texas at Austin, Austin, Texas, United States\\
$^{108}$ Universidad Aut\'{o}noma de Sinaloa, Culiac\'{a}n, Mexico\\
$^{109}$ Universidade de S\~{a}o Paulo (USP), S\~{a}o Paulo, Brazil\\
$^{110}$ Universidade Estadual de Campinas (UNICAMP), Campinas, Brazil\\
$^{111}$ Universidade Federal do ABC, Santo Andre, Brazil\\
$^{112}$ Universitatea Nationala de Stiinta si Tehnologie Politehnica Bucuresti, Bucharest, Romania\\
$^{113}$ University of Derby, Derby, United Kingdom\\
$^{114}$ University of Houston, Houston, Texas, United States\\
$^{115}$ University of Jyv\"{a}skyl\"{a}, Jyv\"{a}skyl\"{a}, Finland\\
$^{116}$ University of Kansas, Lawrence, Kansas, United States\\
$^{117}$ University of Liverpool, Liverpool, United Kingdom\\
$^{118}$ University of Science and Technology of China, Hefei, China\\
$^{119}$ University of South-Eastern Norway, Kongsberg, Norway\\
$^{120}$ University of Tennessee, Knoxville, Tennessee, United States\\
$^{121}$ University of the Witwatersrand, Johannesburg, South Africa\\
$^{122}$ University of Tokyo, Tokyo, Japan\\
$^{123}$ University of Tsukuba, Tsukuba, Japan\\
$^{124}$ Universit\"{a}t M\"{u}nster, Institut f\"{u}r Kernphysik, M\"{u}nster, Germany\\
$^{125}$ Universit\'{e} Clermont Auvergne, CNRS/IN2P3, LPC, Clermont-Ferrand, France\\
$^{126}$ Universit\'{e} de Lyon, CNRS/IN2P3, Institut de Physique des 2 Infinis de Lyon, Lyon, France\\
$^{127}$ Universit\'{e} de Strasbourg, CNRS, IPHC UMR 7178, F-67000 Strasbourg, France, Strasbourg, France\\
$^{128}$ Universit\'{e} Paris-Saclay, Centre d'Etudes de Saclay (CEA), IRFU, D\'{e}partment de Physique Nucl\'{e}aire (DPhN), Saclay, France\\
$^{129}$ Universit\'{e}  Paris-Saclay, CNRS/IN2P3, IJCLab, Orsay, France\\
$^{130}$ Universit\`{a} degli Studi di Foggia, Foggia, Italy\\
$^{131}$ Universit\`{a} del Piemonte Orientale, Vercelli, Italy\\
$^{132}$ Universit\`{a} di Brescia, Brescia, Italy\\
$^{133}$ Variable Energy Cyclotron Centre, Homi Bhabha National Institute, Kolkata, India\\
$^{134}$ Warsaw University of Technology, Warsaw, Poland\\
$^{135}$ Wayne State University, Detroit, Michigan, United States\\
$^{136}$ Yale University, New Haven, Connecticut, United States\\
$^{137}$ Yildiz Technical University, Istanbul, Turkey\\
$^{138}$ Yonsei University, Seoul, Republic of Korea\\
$^{139}$ Affiliated with an institute formerly covered by a cooperation agreement with CERN\\
$^{140}$ Affiliated with an international laboratory covered by a cooperation agreement with CERN.\\

\end{flushleft} 
  
\end{document}